\documentclass[journal]{IEEEtran}

\usepackage[OT1]{fontenc} 
\usepackage[numbers,sort&compress]{natbib}
\usepackage[cmex10]{amsmath}
\usepackage{amssymb}
\usepackage{textcomp}
\usepackage{bm}
\usepackage{braket}
\usepackage{graphicx}
\usepackage{amsthm}
\usepackage{color}
\usepackage{xcolor}
\usepackage{tabularx}
\usepackage{arydshln}
\usepackage{mathtools}
\usepackage{multirow}
\usepackage{algorithm,algorithmic}
\usepackage{url}
\usepackage{listings}
\usepackage{comment}
\usepackage{CJKutf8}
\usepackage[caption=false,font=footnotesize]{subfig} 

\def\CN{\mathcal{CN}}
\def\I{\mathbf{I}}
\def\Y{\mathbf{Y}}

\def\H{\mathbf{H}}

\def\K{\mathbf{K}}
\def\V{\mathbf{V}}

\def\X{\mathbf{X}}

\def\A{\mathbf{A}}

\def\vec{\mathrm{vec}}

\def\P{\mathrm{P}}

\def\U{\mathbf{U}}
\def\C{\mathbb{C}}

\def\x{\mathbf{x}}
\def\u{\mathbf{u}}

\def\w{\mathbf{w}}
\def\E{\mathbb{E}}
\def\F{\mathbf{F}}
\def\S{\mathbf{S}}

\def\F{\mathrm{F}}
\def\e{\mathrm{H}}

\def\G{\mathcal{G}}

\def\P{\mathbf{P}}

\def\Deltab{\mathbf{\Delta}}

\def\kr{\otimes}

\def\K{\mathbf{K}}

\def\col{\text{col}}
\def\O{\mathcal{O}}

\newtheorem{thm}{\textbf{Theorem}}
\newtheorem{lem}{Lemma}
\newtheorem{cor}{Corollary}

\begin{document}
\title{Sparse Grassmannian Design for Noncoherent Codes via Schubert Cell Decomposition}

\author{Joe~Asano,~\IEEEmembership{Student Member,~IEEE}, Yuto Hama,~\IEEEmembership{Member, IEEE}, Hiroki Iimori,~\IEEEmembership{Member, IEEE},\\ Chandan Pradhan,~\IEEEmembership{Member, IEEE}, Szabolcs Malomsoky, and Naoki~Ishikawa,~\IEEEmembership{Senior~Member,~IEEE}.\thanks{J.~Asano and N.~Ishikawa are with the Faculty of Engineering, Yokohama National University, 240-8501 Kanagawa, Japan (e-mail: iskw@ieee.org). Y.~Hama, H.~Iimori, C.~Pradhan, and S.~Malomsoky are with Ericsson Research, Ericsson Japan K. K., Yokohama SYMPHOSTAGE West Tower 12F, 5-1-2 Minato Mirai, Yokohama, 220-0012, Japan (e-mail: [hiroki.iimori, yuto.hama, szabolcs.malomsoky]@ericsson.com).}}

\markboth{\today}
{Shell \MakeLowercase{\textit{et al.}}: Bare Demo of IEEEtran.cls for Journals}
\maketitle

\begin{abstract}
In this paper, we propose a method for designing sparse Grassmannian codes for noncoherent multiple-input multiple-output systems. Conventional pairwise error probability formulations under uncorrelated Rayleigh fading channels fail to account for rank deficiency induced by sparse configurations. We revise these formulations to handle such cases in a unified manner. Furthermore, we derive a closed-form metric that effectively maximizes the noncoherent average mutual information (AMI) at a given signal-to-noise ratio. We focus on the fact that the Schubert cell decomposition of the Grassmann manifold provides a mathematically sparse property, and establish design criteria for sparse noncoherent codes based on our analyses. In numerical results, the proposed sparse noncoherent codes outperform conventional methods in terms of both symbol error rate and AMI, and asymptotically approach the performance of the optimal Grassmannian constellations in the high-signal-to-noise ratio regime. Moreover, they reduce the time and space complexity, which does not scale with the number of transmit antennas.
\end{abstract}


\IEEEpeerreviewmaketitle

\section{Introduction}
\IEEEPARstart{T}{owards} beyond 5G and 6G mobile communication systems, meeting key performance requirements such as ultra-high data rate, ultra-low latency, high connectivity, and cost-efficient implementation is essential~\cite{ngo2025noncoherent,zhang20196g}. In particular, Internet of Things (IoT) networks have been rapidly expanding, accompanied by a significant increase in the number of connected devices. Since many IoT devices primarily generate data, uplink traffic is expected to grow substantially in the future wireless networks~\cite{ericsson2025ericsson}. Consequently, cost-efficient signal processing techniques are required in such large-scale deployments. In this context, reducing computational complexity and improving power efficiency are particularly critical challenges. 

Sparsity is one of the key concepts for addressing such challenges~\cite{ishikawa201850}. It refers to the property that signals or matrices contain only a small number of nonzero elements. By exploiting this property, sparsity can significantly reduce computational complexity and improve power efficiency in signal processing algorithms. In wireless communications, sparsity has been widely utilized in various scenarios, including channel estimation based on compressive sensing~\cite{choi2017compressed}, beamforming design in millimeter-wave channels using orthogonal matching pursuit~\cite{ayach2014spatially}, sparse code multiple access~\cite{yu2021sparse}, and index modulation~\cite{ishikawa2016subcarrierindex}. Moreover, the effectiveness of sparsity in IoT network design has been theoretically demonstrated in~\cite{qin2018sparse}. As next-generation wireless communication systems are expected to involve a massive increase in low-cost devices, further applications and extensions of sparsity-based techniques are anticipated. 

In multiple-input multiple-output (MIMO) systems, rapid channel variations require frequent transmission of reference signals for channel estimation, which increases overhead and reduces transmission efficiency. This problem is particularly pronounced for low-cost devices with limited processing capability and in high-mobility environments, such as bullet trains, where rapid channel variations can render channel estimation highly unreliable.
To address these issues, noncoherent detection has been proposed as a communication technique that does not require channel state information at the receiver. 
In noncoherent communications, differential space-time block codes based on unitary matrices~\cite{hochwald2000differential, ishikawa2017rectangular} have been extensively studied. In these systems, information is differentially precoded across consecutive transmission blocks and decoded without explicit channel estimation by exploiting temporal correlations, at the cost of approximately $3~\text{dB}$ differential loss, because noise from two consecutive blocks contributes to the detection of each symbol. In contrast, using matrices on the Grassmann manifold as transmit signals has been shown to enable robust and efficient noncoherent detection without relying on differential encoding~\cite{hochwald2000unitary, gohary2009noncoherent, gohary2019noncoherent}. 

Various design approaches exist for Grassmannian signaling, including methods that exploit the geometric structure of the Grassmann manifold~\cite{kammoun2003new, kammoun2007noncoherent, ngo2020cubesplit, cuevas2024constellations, cuevas2022measure} and those based on numerical optimization on the manifold~\cite{gohary2009noncoherent, ngo2022joint, cuevas2023union, ngo2020multi-user}. In particular, geometry-based approaches enable not only simple code constructions without optimization but also low-complexity detection algorithms and efficient bit labeling. However, most existing geometric designs are fundamentally based on single-input multiple-output (SIMO) systems. Although efficient code designs and low-complexity detection schemes for MIMO scenarios have been proposed in the literature such as \cite{gohary2009noncoherent}, they are not optimal in terms of the tradeoff between system complexity and performance. Consequently, designing noncoherent codes that effectively balance performance and detection complexity for MIMO systems remains a major challenge.

Furthermore, in~\cite{kammoun2007noncoherent, gohary2009noncoherent, cuevas2023union}, explicit pairwise error probability (PEP) formulations for generalized likelihood ratio test (GLRT) detection under uncorrelated Rayleigh fading channels have been derived and shown to serve as upper bounds on the symbol error rate (SER) for general constellations. However, these formulations do not account for the loss of diversity gain caused by rank deficiency in the underlying subspaces, which often arises from sparse configurations. Thus, it is necessary to formulate a PEP expression that remains valid in such scenarios and accurately characterizes the resulting SER degradation. In addition, although the average mutual information (AMI) under the same detection scheme has been formulated~\cite{endo2024boosting}, a closed-form metric is established that maximizes the AMI for arbitrary constellations, along with corresponding design criteria.

Against this background, we focus on the design of low-complexity and robust noncoherent codes, and we propose the design of a sparse Grassmannian constellation in this paper. The Schubert decompositions of the Grassmann manifold divide the manifold into a finite number of cells characterized by staircase structures~\cite{hansen2007schubert}, thereby inducing inherent sparsity with few nonzero elements. While existing work has proposed noncoherent codes design~\cite{konishi2022novel} based on Schubert decompositions for multi-resolution MIMO systems~\cite{elmossallamy2019multiresolution}, 
\cite{konishi2022novel} poses a challenge that its application in general MIMO scenarios is difficult due to the constraint of sparse configuration. Thus, our objective is to develop sparse constellation designs that are applicable to general noncoherent MIMO communication systems.

The contributions of this paper are summarized as follows:
\begin{enumerate}
    \item We exploit the fact that the Schubert decomposition of the Grassmann manifold induces mathematically sparse structures and propose a sparse Grassmannian constellation for noncoherent MIMO systems. Owing to its inherent sparsity, the proposed constellation enables detection with linear computational complexity in both time and space with respect to the cardinality. 
    
    \item To accurately characterize the SER degradation caused by sparse configurations, we formulate a new PEP expression for GLRT detection under uncorrelated Rayleigh fading channels. We show that the proposed PEP provides valid upper bounds on the SER, consistent with conventional PEP formulations.
    
    \item We derive a closed-form metric that approximately maximizes the noncoherent AMI under uncorrelated Rayleigh fading channels and establish corresponding design criteria for arbitrary constellations. Based on the proposed PEP and AMI analyses, the proposed constellation outperforms conventional designs in terms of both SER and AMI under sparsity constraints, and its performance asymptotically approaches that of theoretically optimal constellations for certain system parameters.
\end{enumerate}

\indent The remainder of this paper is organized as follows. Section~\ref{sec:grassmann} reviews the fundamentals of the Grassmann manifold and conventional constellations. Section~\ref{sec:sys} presents the system model, and Section~\ref{sec:analysis} analyzes general noncoherent codes design criteria in terms of PEP and AMI. Section~\ref{sec:prop} introduces the proposed sparse Grassmannian code design and Section~\ref{sec:res} provides numerical results. Finally, conclusions are drawn in Section~\ref{sec:conc}.

\section{Grassmannian Constellation}
\label{sec:grassmann}
Here, we describe the mathematical definition of the Grassmann manifold and its properties as used in wireless communications.

\subsection{Definition~\cite{bendokat2024grassmann}}
For a positive integer $M$, the unitary group $\mathcal{U}(M)$ is defined as the set of unitary matrices given by
\begin{align}
    \mathcal{U}(M)=\left\{\U\in \C^{M\times M} ~ | ~\U^{\e}\U=\I_{{M}} \right\}.
    \label{eq:unitary}
\end{align}
For a given natural number $T>M$, the Stiefel manifold $\mathcal{S}(T,M)$, which is the set of orthonormal matrices, can be defined as
\begin{align}
    \mathcal{S}(T,M)=\left\{\S \in \C^{T\times M} ~ | ~\S^{\e}\S=\I_{{M}} \right\}
    \label{eq:stiefel}.
\end{align}
Here, let $\mathbf{s}_m$ denote the $m$-th column vector of the matrix $\S$ on the Stiefel manifold $\mathcal{S}(T,M)$. The linear subspace spanned by $\S$ is expressed as
\begin{align}
    \operatorname{span}(\S)=\left\{h_1\mathbf{s}_1+\cdots+h_{{M}}\mathbf{s}_{{M}}~|~h_1,\cdots,h_{{M}}\in\C \right\}
    \label{eq:span}.
\end{align}
At this point, we define the relationship where the linear subspaces spanned by any two points $\S_1$ and $\S_2$ are identical, i.e., $\text{span}(\S_1)=\text{span}(\S_2)$, as equivalence class, and denote it as $\S_1\sim\S_2$. We further denote a set on the Stiefel manifold satisfying such an equivalence class as
\begin{align}
    [\S]=\left\{\S_1,\S_2\in\mathcal{S}(T,M)~|~\S_1\sim\S_2 \right\}
\label{eq:equivalence}.
\end{align}
Here, the Grassmann manifold $\G(T,M)$ is defined as the set of equivalence classes on the Stiefel manifold as
\begin{align}
    \mathcal{G}(T,M)=\mathcal{S}(T,M)/\mathcal{U}(M)=\left\{ [\S]~|~\S\in\mathcal{S}(T,M) \right\}
    \label{eq:grassmann},
\end{align}
that is, the Grassmann manifold $\G(T,M)$ can also be defined as the quotient space by the unitary group $\mathcal{U}(M)$ on the Stiefel manifold $\mathcal{S}(T,M)$.

The Grassmann manifold possesses the property of universality with respect to the operation of multiplication of unitary matrices from right. That is, for any matrix $\X$ on the Grassmannian, the matrix $\X\U$ obtained by multiplying $\X$ from the right by any unitary matrix $\U$ represents exactly the same point on the Grassmann manifold.
As an extension of this property, in~\cite{endo2024boosting, kato2025maximizing}, data carrying reference signal is proposed that enables both channel estimation and information transmission by utilizing Grassmann constellation. Furthermore, the Grassmann manifold is also applied to precoding codebook design. In~\cite{love2003grassmannian, love2005limited, love2008overview}, it is demonstrated that the optimal precoding codebook design in limited feedback system is achieved through maximization of the distances on the Grassmann manifold.

For a given matrix $\X\in\G(T,M)$, let $\mathcal{X}=\left\{\X_1,\cdots,\X_{|\mathcal{X}|} \right\}$ be a Grassmannian constellation with its cardinality $|\mathcal{X}|$. Then, the chordal distance between two points $\X_i$ and $\X_j$, which is denoted as $d_{\mathrm{c}}(\X_i,\X_j)$ is defined as
\begin{align}
\begin{split}
    d_{\mathrm{c}}(\X_i,\X_j)&=\frac{\|\X_i\X_i^\e-\X_j\X_j^\e\|_\F} {\sqrt{2}} \\
    &=\sqrt{M-\|\X_i^\e\X_j\|_\F^2} 
    \label{eq:dch}.
\end{split}
\end{align}
\eqref{eq:dch} can also be denoted using the principal angles defined in~\cite{conway1996packing} as
\begin{align}
    d_\mathrm{c}(\X_i,\X_j)= \sqrt{\sum_{m=1}^{M}\sin^2\theta_{m}^{(i,j)} }
    \label{eq:dch_principal},
\end{align}
where the principal angles $\theta_{1}^{(i,j)},\cdots,\theta_{M}^{(i,j)}$ between $\X_i$ and $\X_j$ are obtained by performing singular value decomposition of the inner product matrix $\X_i^\e\X_j$, which is denoted as
\begin{align}
    \X_i^\e\X_j=\mathbf{U}&\operatorname{diag}(\sigma_{1}^{(i,j)},\cdots,\sigma_{M}^{(i,j)})\mathbf{V}^\e
    \label{eq:SVD}.
\end{align}
In~\eqref{eq:SVD}, $\sigma_{1}^{(i,j)}\geq \cdots\geq\sigma_{M}^{(i,j)}\geq0$ are singular values, and $\U,\V\in\mathcal{U}(M)$. Using the singular values obtained from~\eqref{eq:SVD}, the principal angles $\theta_{1}^{(i,j)},\cdots,\theta_{M}^{(i,j)}$ are defined as
\begin{align}
    \theta_{m}^{(i,j)}=\arccos(\sigma_{m}^{(i,j)}),~~m=1,\cdots,M
    \label{eq:p_angle}.
\end{align}
In this paper, let the projection matrix corresponding to $\X$ be defined as $\P = \X\X^\e$. That is, we denote the projection matrix of $\X_i$ as $\P_i$.

\subsection{Conventional Grassmannian Constellation}
We review two conventional Grassmannian constellations that support multiple transmit antennas $M \geq 2$, since our proposed method is developed for such scenarios and these methods serve as appropriate benchmarks. Many appealing Grassmannian constellations have been designed for SIMO settings, i.e., $M=1$.
A comprehensive review can be found in \cite{ngo2025noncoherent}.

\subsubsection{Manopt}
The constellation design method based on direct optimization over the Grassmann manifold is referred to as Manopt in this paper. The method that maximizes the minimum chordal distance (MCD), which is defined as
\begin{align}
    d_{\mathrm{min}} = \underset{1\leq i<j\leq |\mathcal{X}|} {\operatorname{min}} d_\mathrm{c}(\X_i,\X_j),
\end{align}
through optimization using (\ref{eq:dch}) and (\ref{eq:dch_principal}) is referred to as MCD-Manopt. The MCD maximization problem can be formulated as
\begin{align}
    \underset{\mathcal{X}}{\operatorname{maximize}}& \underset{1\leq i<j\leq |\mathcal{X}|} {\operatorname{min}} d_{\mathrm{c}}(\X_i,\X_j)
    \label{eq:obj_MCD}\\
    \text{s.t.}&~~\X_i\in\mathcal{G}(T,M),~\forall ~i=\left\{1,\cdots,|\mathcal{X}| \right\}, \nonumber
\end{align}
where the objective function in (\ref{eq:obj_MCD}) can be approximated by a smooth surrogate objective as
\begin{align}
    \underset{\mathcal{X}}{\operatorname{minimize}}~~&\log \underset{1\leq i<j\leq |\mathcal{X}|}{\sum}\exp\left(-\frac{\|\X_i\X_i^\e-\X_j\X_j^\e\|_\F}{\epsilon} \right)
    \label{eq:obj_MCD_2}\\
    \text{s.t.}~~&\X_i\in\mathcal{G}(T,M),~\forall ~i=\left\{1,\cdots,|\mathcal{X}| \right\}, \nonumber
\end{align}
with $\epsilon$ a smoothing constant.

\subsubsection{Exp-Map~\cite{kammoun2003new, kammoun2007noncoherent}}
Exponential map (Exp-Map) is a standard map that projects complex matrices onto points on the Grassmann manifold.
Using the matrix exponential, we can represent any point on the Grassmann manifold as
\begin{align}
    \X=\left[ \exp
    \begin{pmatrix}
        \mathbf{0} & \mathbf{\Theta} \\
        -\mathbf{\Theta}^\e & \mathbf{0} 
    \end{pmatrix}
     \right] \I_{T,M},
     \label{eq:exp}
\end{align}
where $\I_{T,M}=[\I_M ~\mathbf{0}_{M\times(T-M)}]^\mathrm{T}$ and $\mathbf{\Theta}\in\C^{M\times(T-M)}$ is an arbitrary matrix, such as space-time block codes composed of quadrature amplitude modulation (QAM) symbols. When $M=1$, $T-1$ symbols are simply mapped to each element of $\mathbf{\Theta}$.
The advantage of Exp-map is that it supports MIMO scenarios, i.e., $M\geq2$. 
In such cases, coherent space-time codes are used~\cite{kammoun2007noncoherent}. For instance, when $(T,M)=(4,2)$, the matrix $\mathbf{\Theta}$ is constructed as
\begin{align}
    \mathbf{\Theta} = \begin{bmatrix}
        s_1+\theta s_2 & \phi(s_3+\theta s_4) \\
        \phi (s_3-\theta s_4) & s_1-\theta s_2
    \end{bmatrix}
    \label{eq:Theta},
\end{align}
where $s_1,\cdots,s_4$ are 4-QAM symbols and the parameters are $\phi^2=\theta=e^{j\pi/4}$.

\section{System Model}
\label{sec:sys}
We consider a system with $M$ transmit antennas and $N$ receive antennas as well as the identically distributed (i.i.d.) Rayleigh fading channels $\H\in\C^{M\times N}$, which remain constant over $T$ time slots and follow Gaussian distribution $\CN(0,1)$.
A space-time block code (STBC) $\S=\sqrt{T/M}\X\in\C^{T\times M}$ satisfies the power constraint $\mathbb{E}[\|\S\|_\F^2]=\E[\|\sqrt{T/M}\X\|_\F^2]=T$. 
Here, let $\mathcal{X}=\left\{\X_1,\cdots,\X_{|\mathcal{X}|}  \right\}$ be a Grassmannian constellation with cardinality $|\mathcal{X}|$. Then, when each code $\X \in \C^{T \times M}$ is a point on the Grassmann manifold, the received signal $\Y\in\C^{T\times N}$ is given by
\begin{align}
    \Y=\S\H+\V=\sqrt{\frac{T}{M}}\X\H+\V,
    \label{eq;sys_noncoherent}
\end{align}
where $\V \in \C^{T \times N}$ is the additive noise matrix with i.i.d. entries $\CN(0, \sigma_v^2)$, and the SNR is defined as $\mathrm{SNR} = 10\log_{10}(1/\sigma_v^2)$ [dB]. At the receiver, the GLRT detector estimates $\hat{\X}$ as
\begin{align}
    \hat{\X} = \underset{\X\in\mathcal{X}}{\operatorname{argmax}} ~\operatorname{tr}(\Y\Y^\e\X\X^\e),
    \label{eq:GLRT}
\end{align}
where $\operatorname{tr}(\cdot)$ denotes the trace. Assuming this detection scheme, its noncoherent AMI is given by \cite{endo2024boosting}
\begin{align}
     R_g= \frac{\log_2|\mathcal{X}|}{T}-\frac{1}{T|\mathcal{X}|}\E_{\H,\V}\left[\sum_{i=1}^{|\mathcal{X}|}\log_2\sum_{j=1}^{|\mathcal{X}|}\exp(\eta_{i,j})\right],
     \label{eq:AMI}
\end{align}
where $\eta_{i,j}$ is defined as
\begin{align}
     \eta_{i,j} = \frac{-\operatorname{tr}(\Y_i\Y_i^\e\Deltab)}{\sigma_v^2(1+\sigma_v^2\frac{M}{T})} 
     \label{eq:eta},
\end{align}
with $\Deltab=\X_i\X_i^\e-\X_j\X_j^\e$.

\section{Mathematical Analysis}
\label{sec:analysis}
In this section, we describe the fundamental analysis for noncoherent detection in terms of PEP and AMI.
First, we modify the formulation of PEP for GLRT detection considering the rank-deficient case, which is a novel contribution of this paper.
Next, we analyze the noncoherent AMI defined in \eqref{eq:AMI} and introduce the metric that maximizes it.  

\subsection{PEP Analysis for General Dense and Sparse Codes}
PEP is the probability of detecting $\X_j$ incorrectly when $\X_i$ is transmitted, and for GLRT detection (\ref{eq:GLRT}) it is given by \cite{kammoun2007noncoherent} 
\begin{align}
    P(\X_i\rightarrow \X_j) &\leq \frac{\sigma_v^{2MN}\binom{2MN-1}{MN}}{\operatorname{Re}[\operatorname{det}(\I_M-\X^\e_i\X_j\X^\e_j\X_i)]^N}
     \label{eq:pep_conv},
\end{align}
Here, when designing the sparse constellation, rank-deficient problems arise with conventional criteria \cite{cuevas2023union,ngo2022joint,alvarez-vizoso2023constrained}. Specifically, when the matrix $\I_M - \X^\e_i\X_j\X^\e_j\X_i$ is rank-deficient, some of its eigenvalues become zero, causing the determinant to vanish. As a result, the denominator of (\ref{eq:pep_conv}) becomes zero, making the expression undefined or divergent. Thus, this expression is inadequate for cases in which $\I_M - \X^\e_i\X_j\X^\e_j\X_i$ is not of full rank.

To resolve this issue, we propose a revised upper bound on the PEP that accounts for rank deficiency induced by sparse codes. The proposed PEP expression under the same channel assumption is given as
 \begin{align}
    {P}(\X_i\rightarrow \X_j) &\leq \frac{\sigma_v^{2m'N}\binom{2m
    'N-1}{m'N}}{\prod_{m=1}^{m'}|\mu_m|^N}
    \label{eq:PEP_propsed},
\end{align}
where $1\le m'\le M$ is the rank of $\I_M-\X^\e_i\X_j\X^\e_j\X_i$, and each $\mu_m$ denotes the nonzero eigenvalue of $\I_M-\X^\e_i\X_j\X^\e_j\X_i$. This form is commonly seen in the context of coherent STBC.

Although \eqref{eq:PEP_propsed} is a minor modification of~\eqref{eq:pep_conv} proposed in~\cite{kammoun2007noncoherent}, it provides new insights and this formulation is applicable for general dense constellations. 
We see that the diversity gain is maximized when $m'=M$, which makes $\I_M-\X^\e_i\X_j\X^\e_j\X_i$ full rank.
The denominator of (\ref{eq:pep_conv}) is defined as the chordal product distance. For two points $\X_i$ and $\X_j$, its chordal product distance $d_{\mathrm{cp}}(\X_i,\X_j)$ is defined as \cite{kammoun2007noncoherent, alvarez-vizoso2022statistical}
\begin{align}
    d_{\mathrm{cp}}(\X_i,\X_j) &=\det\left( \I_M-\X^\e_i\X_j\X^\e_j\X_i \right).
    \label{eq:dcp}
\end{align}
By using the principal angles denoted in~\eqref{eq:p_angle}, $d_{\mathrm{cp}}(\X_i,\X_j)$ can also be denoted as
\begin{align}
    d_{\mathrm{cp}}(\X_i,\X_j) = \prod_{m=1}^{M}\sin^2\theta_{m}^{(i,j)}.
    \label{eq:dcp_principal}
\end{align}
In particular, when $M=1$, there is only one principal angle between any pair of two points and the chordal product distance (\ref{eq:dcp_principal}) becomes identical to chordal distance \eqref{eq:dch_principal}.
Hence, this insight clarifies that it is important to maximize the minimum chordal product distance (MCPD) to minimize the SER. We employ the union bound of the PEP as an upper bound on the SER. For a given constellation $\mathcal{X}$, the union bound is given by \cite{gohary2009noncoherent}  
\begin{align}
      {P_U} &= \frac{2}{|\mathcal{X}|}\sum_{i=1}^{|\mathcal{X}|-1}\sum_{j=i+1}^{|\mathcal{X}|}{P}(\X_i\rightarrow \X_j).
      \label{eq:union_bound}
  \end{align}

\subsection{Closed-Form Lower Bound for AMI}
We derive a closed-form metric that effectively maximizes~\eqref{eq:AMI}.
Since $\exp(\eta_{i,j})>0$, the lower bound of the AMI given in~\eqref{eq:AMI} can be expressed using Jensen-type inequality as
\begin{align}
    R_g
    &\ge \frac{\log_2|\mathcal{X}|}{T}-\frac{1}{T|\mathcal{X}|\lambda\ln2}\sum_{i=1}^{|\mathcal{X}|}\ln\sum_{j=1}^{|\mathcal{X}|}\E_{\H,\V}\left[ \exp(\lambda\eta_{i,j}) \right],
    \label{ineq:AMI}
\end{align}
where $\lambda$ is a concavity parameter that satisfies $0<\lambda\le1$. 
The metric that approximately maximizes~\eqref{ineq:AMI} is given by the following theorem.
\begin{thm}
\label{thm:conc}
When the principal angles between $\X_i$ and $\X_j$ are defined as~\eqref{eq:p_angle}, the lower bound of the AMI given in~\eqref{ineq:AMI} can be expressed as
\begin{align}
    R_g
        &\ge \frac{\log_2|\mathcal{X}|}{T}-\frac{1}{T|\mathcal{X}|\lambda\ln2}\sum_{i=1}^{|\mathcal{X}|}\ln\sum_{j=1}^{|\mathcal{X}|}\mathcal{E}_{i,j}
    \label{eq:conc}
\end{align}
where
\begin{align}
    \mathcal{E}_{i,j}=\left[\prod_{m=1}^{M} {\left(1+\kappa(\lambda)\sin^2\theta_m^{(i,j)}\right)} \right]^{-N},
    \label{eq:mathcal_E}
\end{align}
with
\begin{align}
    \kappa(\lambda)=\alpha-\alpha^2\sigma_v^2-\alpha^2\sigma_v^4
    \label{eq:kappa_s},
\end{align}
and
\begin{align}
 \alpha= \frac{\lambda}{\sigma_v^2\left(1+\sigma_v^2\frac{M}{T}\right)}.
 \label{eq:alpha}
\end{align}
\end{thm}

\begin{proof}
    Refer to Appendix~\ref{app:conc}.
\end{proof}

For a given SNR, the noise variance $\sigma_v^2$ is fixed and the concavity parameter $\lambda$ should be adjusted to tighten the Jensen-type bound in~(\ref{ineq:AMI}). Since $\kappa(\lambda)$ is a concave quadratic function of $\lambda$, the optimal value of $\lambda$ that maximizes $\kappa(\lambda)$ thus minimizing $\mathcal{E}_{i,j}$ is obtained in closed form as follows.

\begin{thm}
\label{thm:lambda_star}
For a given $\mathrm{SNR} = 1/\sigma_v^2$, the parameter $\lambda$ that tightens the Jensen-type lower bound given in~\eqref{ineq:AMI} is obtained in closed-form as
    \begin{align}
        \lambda^{\star}=\min\left\{1, \frac{1+\sigma_v^2M/T}{2(1+\sigma_v^2)}\right\}
        \label{eq:s_star}.
    \end{align}
When $\lambda=\lambda^\star$, the function $\kappa(\lambda^\star)$ increases monotonically with respect to $\mathrm{SNR}$ and $\kappa(\lambda^\star)=1$ if and only if $\mathrm{SNR}=6.84~\mathrm{dB}$, independently of $T$ and $M$.
\end{thm}

\begin{proof}
    The expression in (\ref{eq:s_star}) is obtained by maximizing the concave quadratic $\kappa(\lambda)$ over the interval $0<\lambda\le 1$, resulting in a clipped quadratic maximizer. Substituting $\lambda^\star$ into $\kappa(\lambda)$ shows that
    $\kappa(\lambda^\star)$ increases monotonically with SNR, and solving $\kappa(\lambda^\star)=1$ leads to the SNR threshold of $6.84~\mathrm{dB}$.
\end{proof}

\begin{figure}[tb]
    \centering
    \includegraphics[scale=0.7]{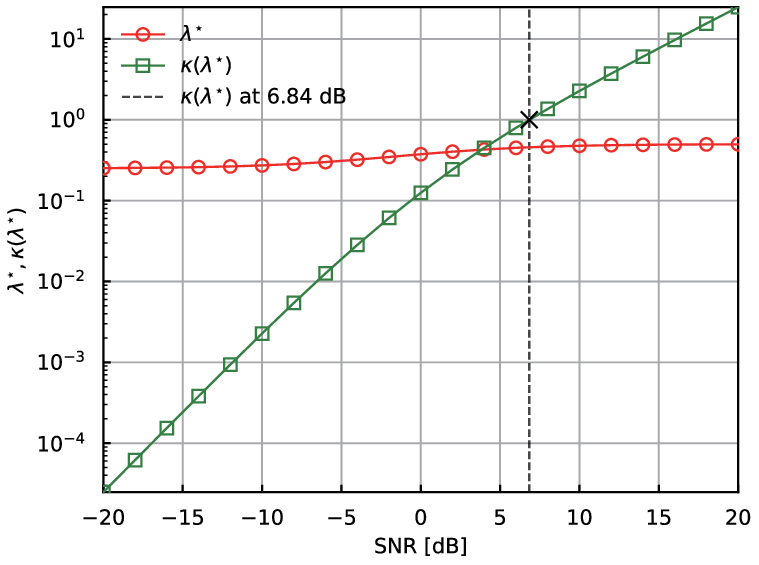}
    \caption{Variations in $\lambda^\star$ and $\kappa(\lambda^\star)$ with respect to SNR when $T/M=2$.}
    \label{fig:kappa_s}
\end{figure}

Next, we investigate the behavior of the dominant term in~\eqref{eq:mathcal_E} with respect to a given SNR. For this purpose,
we focus on the term $\prod_{m=1}^{M}(1+\kappa(\lambda^\star)\sin^2\theta_m^{(i,j)})$ in~\eqref{eq:mathcal_E}. For example, when $M=2$, this product can be expanded to a sum of four terms, which is denoted as
\begin{align}
    \prod_{m=1}^{2} \big(1+&\kappa(\lambda^\star)\sin^2\theta_m^{(i,j)}\big) \notag \\
    =1+&\kappa(\lambda^\star)\big(\sin^2\theta_1^{(i,j)}+\sin^2\theta_2^{(i,j)}\big) \notag \\
    +&\kappa(\lambda^\star)^2\sin^2\theta_1^{(i,j)}\sin^2\theta_2^{(i,j)}
\label{eq:ks_decomposed}.
\end{align}
By combining this expansion with~\eqref{eq:dch_principal} and~\eqref{eq:dcp_principal}, the resulting expression can be interpreted as a linear combination of the chordal distance and the chordal product distance, where each component is weighted by powers of $\kappa(\lambda^\star)$. Based on this observation, we refer to this combined measure as a joint distance. The joint distance between two points $\X_i$ and $\X_j$, denoted by $d_{\text{joint}}(\X_i,\X_j)$, is defined as
\begin{align}
    &1+\kappa(\lambda^\star)\underbrace{(\sin^2\theta_1^{(i,j)}+\sin^2\theta_2^{(i,j)})}_{\text{chordal distance}} \notag \\
    &~~~~~~~~~~~~~~~~~~~~~~~~~~~~~~~~~+\kappa(\lambda^\star)^2\underbrace{\sin^2\theta_1^{(i,j)}\sin^2\theta_2^{(i,j)}}_{\text{chordal product distance}} \notag \\
    =&1+\kappa(\lambda^\star)d_{\mathrm{c}}(\X_i,\X_j)^2+\kappa(\lambda^\star)^2d_{\mathrm{cp}}(\X_i,\X_j) \notag \\
    :=&d_{\text{joint}}(\X_i,\X_j)
    \label{eq:d_joint}.
\end{align}
The same property holds for arbitrary $M$, and the joint distance $d_{\text{joint}}(\X_i,\X_j)$ is simply given by 
\begin{align}
    d_{\text{joint}}(\X_i,\X_j)=\prod_{m=1}^{M} \left({1+\kappa(s^\star)\sin^2\theta_m^{(i,j)}}\right).
\end{align}

In conjunction with the result of~\eqref{ineq:AMI}, the objective function for approximately maximizing AMI can be expressed as the maximization of the sum of the joint distances over all distinct pairs of constellations, which is denoted as
\begin{align}
    \underset{\mathcal{X}}{\operatorname{maximize}}~~\sum_{1\le i<j\le|\mathcal{X}|}d_{\text{joint}}(\X_i,\X_j)
    \label{eq:obj_AMI}.
\end{align}

\begin{cor}
\label{cor:d_joint}
Let $\lambda^\star$ be given by Theorem~\ref{thm:lambda_star}. The objective function in \eqref{eq:obj_AMI} for maximizing AMI can be formally approximated as
\begin{align}
&\underset{\mathcal{X}}{\operatorname{maximize}}~~\sum_{1\le i<j\le|\mathcal{X}|}d_{\mathrm{joint}}(\X_i,\X_j)  \notag \\
    \simeq&\left\{\begin{array}{cc}
        \underset{\mathcal{X}}{\operatorname{maximize}}~\sum~d_{\mathrm{cp}}(\X_i,\X_j) & (\mathrm{SNR}>6.84~\mathrm{dB}) \\
        \underset{\mathcal{X}}{\operatorname{maximize}}~\sum~d_{\mathrm{c}}(\X_i,\X_j) & (\mathrm{SNR}\le6.84~\mathrm{dB})
    \end{array} \right.
    \label{eq:obj_AMI_2}.
\end{align}
\end{cor}

\begin{proof}
    This follows directly from the monotonic increase of $\kappa(\lambda^\star) $ with SNR and the fact that $\kappa(\lambda^\star)=1$ exactly at $6.84~\mathrm{dB}$, which determines the switching point between chordal and chordal product distance behavior. 
\end{proof}

In~\cite{kato2025maximizing}, with $\text{SNR}=0~\text{dB}$, the maximization of AMI can be approximated by maximizing the sum of the chordal distance among the constellation. The result in~\eqref{eq:obj_AMI_2} is consistent with the observation claimed by~\cite{kato2025maximizing}. 
However, since it is difficult to employ the optimal parameter $\lambda^\star$ for an arbitrary noise variance $\sigma_v^2$, we instead consider a high-SNR approximation while taking into account the trade-off with error rate minimization. With respect to~\eqref{eq:obj_AMI_2}, in the high-SNR regime, the problem of maximizing the AMI can be approximated by maximizing the chordal product distance among the constellation points.
Accordingly, we propose a code design method, referred to as MCPD-Manopt, which aims to maximize the MCPD between codewords. The corresponding optimization problem is formulated as
 \begin{align}
    \underset{\mathcal{X}}{\operatorname{minimize}}~~&\log \underset{1\leq i<j\leq |\mathcal{X}|}{\sum}\exp\left(-\frac{\det(\I_M-\X_i^\e\X_j\X_j^\e\X_i)}{\epsilon} \right)
    \label{eq:obj_MCPD}\\
    \text{s.t.}~~&\X_i\in\mathcal{G}(T,M),~\forall ~i=\left\{1,\cdots,|\mathcal{X}| \right\}, \nonumber
\end{align}
 with $\epsilon$ a smoothing constant.

\section{Proposed Design Method}
\label{sec:prop}
In this section, we describe the design method of proposed sparse Grassmannian constellation based on the analyses presented in Section~\ref{sec:analysis}.

\subsection{General Design Method}
The general design of proposed sparse Grassmann codes can be summarized for given $T>M>1$ and cardinality $|\mathcal{X}|$ as follows:
\begin{enumerate}
    \item Partition the Grassmann manifold $\mathcal{G}(T,M)$ into disjoint groups using the Schubert cell decomposition. For each partitioned Schubert cell, define sparsity patterns satisfying the column-orthogonality condition.
    \item Select the sparsity patterns corresponding to the desired cardinality $|\mathcal{X}|$, and embed an arbitrary complex number $\ast$ into the nonzero elements of the selected sparsity pattern. At this point, ensure $\I_M-\X_i^\e\X_j\X_j^\e\X_i$ achieves full-rank for all matrices $i\neq j$. 
    \item Maximize the MCPD in the constellation.
\end{enumerate}
Note that the proposed method is applicable to any $T$ and $M$ that satisfy $T > M > 1$, i.e., only MIMO scenarios. We explain each step in detail as follows.

\subsubsection{Schubert Cell Decomposition \cite{konishi2022novel,hansen2007schubert}}
The complex Grassmann manifold $\mathcal{G}(T,M)$ can be decomposed into a disjoint union of Schubert cells
\begin{align}
    \mathcal{G}(T,M)=\bigsqcup_{\mathbf{k}\in[T]}\mathbf{C}_{\mathbf{k}},
    \label{eq:Schubert_decompose}
\end{align}
where each index $\mathbf{k} = \{k_1,\cdots,k_M\}$ defines a Schubert cell $\mathbf{C}_{\mathbf{k}}$, and the symbol $\bigsqcup$ denotes a disjoint union.
Each cell $\mathbf{C}_{\mathbf{k}}$ can be uniquely represented by a $T\times M$ matrix in column echelon form, such that the $(k_l,l)$-th entry is $1$ for $l=1,\cdots,M$, with all entries above that position in the same column and all entries to the left in the same row equal to zero. The other entries to the right in the same row or below in the same column may take arbitrary complex values. In the subsequent discussion in this paper, we denote this arbitrary complex number as $\ast$.

\subsubsection{Lifting Schubert Cells onto the Complex Grassmann}
For a given matrix, the number of its nonzero elements is defined as $s$ and referred to as the sparsity.
For a given sparsity $s$, we further refine this decomposition by considering the sparsity patterns, i.e., the combinations of nonzero entries within the matrices representing each Schubert cell, that yield column-orthogonal configurations, thereby representing points on the complex Grassmann manifold. 
In this paper, in order to ensure sparsity as well as column orthogonality in a straightforward manner, we impose the following constraint, that is, 
\begin{itemize}
    \item Sparsity constraint: $M\le s\le T$
    \item Column-orthogonal constraint: the supports of the nonzero entries in different columns are mutually disjoint.
\end{itemize}
In other words, no row index is shared between two different columns. Under this condition, the columns have disjoint supports, so their inner products automatically vanish. As a result, any such matrix is sparse and column orthogonal, i.e., on the complex Grassmann manifold without further adjustment. Moreover, under this constraint, the number of admissible sparsity patterns can be counted easily, one first chooses $s$ rows out of $T$, and then partitions them into $M$ nonempty disjoint subsets corresponding to the column supports. Thus, for a given $T,M$ and sparsity $s$, its total number of sparsity patterns $n(T,M,s)$ is given by
\begin{align}
    n(T,M,s)=\frac{\binom{T}{s}}{M!}\sum_{k=0}^{M}(-1)^k\binom{M}{k}(M-k)^s
    \label{eq:n(T,M,s)}.
\end{align}
Next, matrices are selected from the obtained $n(T,M,s)$ sparsity patterns. Arbitrary complex values are then embedded to the nonzero elements denoted by $\ast$, and the resulting matrices are normalized before performing optimization.

At this point, to minimize SER, for the selected sparsity pattern, we must care to ensure that $\I_M-\X_i^\e\X_j\X_j^\e\X_i$ achieves full rank, as shown in \eqref{eq:PEP_propsed}. That is, when a sparsity pattern allows only one nonzero element in any given column, selecting two or more matrices from that single sparsity pattern reduces the rank of $\I_M-\X_i^\e\X_j\X_j^\e\X_i$, thereby diminishing the representable subspace. We must avoid such sparsity patterns during selection.

Here, let ${p}$ denote the set of complex parameters for a given matrix $\mathbf{X}$. ${p}$ can be defined as the set of parameters for the amplitude and phase of $s$ complex numbers. 
For example, $p$ can be defined as $p=\{\alpha_1,\phi_1,\cdots\}$, where $\alpha_i,\phi_i$ denotes the amplitude and phase of the $i$-th complex elements.
For a given cardinality $\mathcal{X}$, we define the set of these parameters as $\mathcal{P}=\{{p}_1,\cdots,{p}_{|\mathcal{X}|}\}$ and perform optimization with respect to $\mathcal{P}$. Here, each parameter can be freely set to both discrete and continuous values under arbitrary applications, as long as the constraints are satisfied.

\subsubsection{Optimization}
Here, we optimize the constellation based on the parameters defined above.
When we consider designing a constellation $\mathcal{X} = \{\X_1, \cdots, \X_{|\mathcal{X}|}\}$ of size $|\mathcal{X}|$ and obtain a set of parameters $\mathcal{P}=\left\{p_1,\cdots,p_{|\mathcal{X}|} \right\}$, the optimization objective,
\begin{align}
    \underset{\mathcal{P}}{\operatorname{minimize}}~~&\log \underset{1\leq i<j\leq |\mathcal{X}|}{\sum}\exp\left(-\frac{\det\left(\I_M-\X_i^\e\X_j\X_j^\e\X_i\right)}{\epsilon} \right)
    \label{eq:obj_1},\\
    \text{s.t.}~~&\mathcal{P} = 
    \left\{p_1,\cdots,p_{|\mathcal{X}|}\right\}, \notag
\end{align}
where $\epsilon$ is a smoothing constant. Each parameter set $p_i$ in $\mathcal{P}$ is optimized independently under the constraints defined above. 

\subsection{Example Case for $(T,M)=(4,2)$}
Here, we consider the design of noncoherent codes with $T=4,~M=2$, aiming to maximize MCPD based on~\eqref{eq:obj_MCPD}. 
\subsubsection{Schubert Cell Decomposition}
First, $\G(4,2)$ can be decomposed into the following disjoint six Schubert cells, 
\begin{align}
    \begin{bmatrix}
    1 & 0 \\
    0 & 1 \\
    \ast & \ast \\
    \ast & \ast
    \end{bmatrix}
    ,
    \begin{bmatrix}
    1 & 0 \\
    \ast & 0 \\
    0 & 1 \\
    \ast & \ast
    \end{bmatrix}
    ,
    \begin{bmatrix}
    0 & 0 \\
    1 & 0 \\
    0 & 1 \\
    \ast & \ast
    \end{bmatrix}
    ,
    \begin{bmatrix}
    1 & 0 \\
    \ast & 0 \\
    \ast & 0 \\
    0 & 1
    \end{bmatrix}
    ,
    \begin{bmatrix}
    0 & 0 \\
    1 & 0 \\
    \ast & 0 \\
    0 & 1
    \end{bmatrix}
    ,
    \begin{bmatrix}
    0 & 0 \\
    0 & 0 \\
    1 & 0 \\
    0 & 1
    \end{bmatrix},
\label{eq:schubert cells}
\end{align}
where $\ast$ represents an arbitrary complex number. 

\subsubsection{Lifting Schubert Cells onto the Complex Grassmann}
Next, for the decomposed cells, we select the sparsity patterns that satisfy the column-orthogonality constraint. In this case, we consider $s=4$, i.e., $4$-sparse matrices, and obtain $n(4,2,4)=7$ sparsity patterns in this case.
These sparsity patterns are summarized in Table \ref{tab:sparsity_patterns}.

\begin{table}[tb]
  \caption{Examples of Sparsity Patterns for $(T, M) = (4, 2)$ and $s=4$.}
  \label{tab:sparsity_patterns}
  \centering
  \begin{tabular}{ccccc}
    \hline
    \textbf{Indices} & \multicolumn{4}{c}{\textbf{Matrices}} \\ \hline
    1-4 & 
    $
    \begin{bmatrix}
      1 & 0\\
      0 & 1\\
      \ast & 0\\
      0 & \ast
    \end{bmatrix}$ &
    $\begin{bmatrix}
      1 & 0\\
      0 & 1\\
      0 & \ast\\
      \ast & 0
    \end{bmatrix} $&
    $
    \begin{bmatrix}
      1 & 0\\
      0 & 1\\
      0 & \ast\\
      0 & \ast
    \end{bmatrix}$ &
    $\begin{bmatrix}
      1 & 0\\
      0 & 1\\
      \ast  & 0\\
      \ast & 0
    \end{bmatrix}$ \\ \hline

    5-7 & 
    $\begin{bmatrix}
      1 & 0\\
      \ast & 0\\
      0  & 1\\
      \ast & 0
    \end{bmatrix}$ &
    $\begin{bmatrix}
      1 & 0\\
      \ast & 0\\
      0 & 1\\
      0 & \ast
    \end{bmatrix}$ &
    $\begin{bmatrix}
      1 & 0\\
      \ast & 0\\
      \ast & 0\\
      0 & 1
    \end{bmatrix}$
    & 
     \\ \hline
  \end{tabular}
\end{table}
As shown in Table \ref{tab:sparsity_patterns}, regardless of which complex numbers are assigned to each $\ast$, the columns are mutually orthogonal.

Then, we select sparsity patterns from Table \ref{tab:sparsity_patterns} for a given cardinality $|\mathcal{X}|$. As discussed previously, we must avoid selection that results in rank of $\I_M-\X_i^\e\X_j\X_j^\e\X_i$ being deficient. For example, we select the sparsity pattern with index $3$ to construct two matrices $\X_i, \X_j$, that is, 
\begin{align}
    \X_i\in\begin{bmatrix}
      1 & 0\\
      0 & 1\\
      0 & \ast\\
      0 & \ast
    \end{bmatrix},~\X_j\in\begin{bmatrix}
      1 & 0\\
      0 & 1\\
      0 & \ast\\
      0 & \ast
    \end{bmatrix}.
    \notag
\end{align}
Since the first columns of matrices $\X_i$ and $\X_j$ are identical, they share the same subspace spanned by the first column.
Consequently, $\I_M-\X_i^\e\X_j\X_j^\e\X_i$ becomes rank-deficient, i.e., $\text{rank}(\I_M-\X_i^\e\X_j\X_j^\e\X_i)=1$. Similar rank-deficiency behavior occurs in sparsity patterns with index 4, 5, and 7 when they are used to generate multiple distinct matrices from the same pattern. For this reason, these patterns are not suitable for constructing multiple codewords.
On the contrary, sparsity patterns such as indices 1, 2, and 6 allow the two columns to occupy structurally different positions. This enables each matrix to represent distinct column subspaces provided that the embedded complex parameters are selected to avoid coincident columns. In practice, different column-support structures ensure distinct projection behaviors between codewords, which empirically and analytically leads to full-rank $\I_M-\X_i^\e\X_j\X_j^\e\X_i$ in all examined case.
Hence, these sparsity patterns are adopted when multiple matrices are required. When matrices are selected from different sparsity indices, their structural differences in column support make the probability of rank deficiency is negligible under valid parameterization, which is consistent with the underlying disjoint-support framework.

After selecting valid sparsity patterns, complex numbers are embedded into the $\ast$ elements and the matrices are parametrized to ensure that they lie on the Grassmann manifold $\G(4,2)$.
For instance, if the matrix selected from index~1 in Table~\ref{tab:sparsity_patterns} is denoted $\X_1$, the matrix after embedding the complex numbers is given as 
\begin{align}
    \X_1= \begin{bmatrix}
    \sin\alpha_1 & 0\\
      0 & \sin{\alpha_2}\\
      e^{\mathrm{j}\phi_1}\cos\alpha_1 & 0\\
      0 & e^{\mathrm{j}\phi_2}\cos{\alpha_2}
    \end{bmatrix}
    \label{eq:X1},
\end{align}
where $\alpha_i\in[-\pi,\pi]$ represents the angular parameter which controls the normalization of each column, and $\phi_i\in[-\pi,\pi]$ represents the phase parameter of the complex number. Hence, the set of parameters $p_1$ assigned for the matrix $\X_1$ is given by $p_1=\{\alpha_1,\phi_1,\alpha_2,\phi_2\}$. Parameters of this type are independently assigned to each matrix, and the complete parameter set for any cardinality $|\mathcal{X}|$ is expressed as $\mathcal{P}=\{p_1,\cdots,p_{|\mathcal{X}|}\}$.

\subsubsection{Optimization}
Finally, for the set of parameters $\mathcal{P}$ defined with respect to the cardinality $|\mathcal{X}|$, we perform optimization defined in~\eqref{eq:obj_1} as
\begin{align}
        \underset{\mathcal{P}}{\operatorname{minimize}}~~&\log \underset{1\leq i<j\leq |\mathcal{X}|}{\sum}\exp\left(-\frac{\det\left(\I_M-\X_i^\e\X_j\X_j^\e\X_i\right)}{\epsilon} \right)
    \label{eq:obj_2},\\
    \text{s.t.}~~&\mathcal{P} = 
    \left\{p_1,\cdots,p_{4}\right\}. \notag
\end{align}
For example, when $|\mathcal{X}|=4$, four real parameters are assigned to each codeword, resulting in a total of sixteen real optimization parameters, and they are optimized independently.
In this paper, we set a smoothing constant $\epsilon=10^{-2}$.

\section{Performance Comparisons}
\label{sec:res}
In this section, we compare the performance of the proposed sparse noncoherent codes in terms of SER and AMI with the conventional constellations of MCD-Manopt and Exp-Map as well as reference constellations of MCPD-Manopt and the rank-deficient case described in Section~\ref{sec:analysis}.
The numbers of time slots and transmit antennas were set to $(T,M)=(4,2)$, $(6,2)$, and $(6,3)$. The number of receive antennas $N$ was set to $M$, and the sparsity, i.e., the number of nonzero elements $s$ was set to $T$ for each setting. We also evaluate the scalability of the proposed method with respect to the cardinality $|\mathcal{X}|$ and discuss the computational-complexity reduction achieved by the proposed method.

\subsection{SER Comparison}
First, we compare SER performance and evaluate the proposed PEP formulation with the detection based on the GLRT detector defined in~\eqref{eq:GLRT}.

\begin{figure}[t]
    \centering
    \includegraphics[scale=0.7]{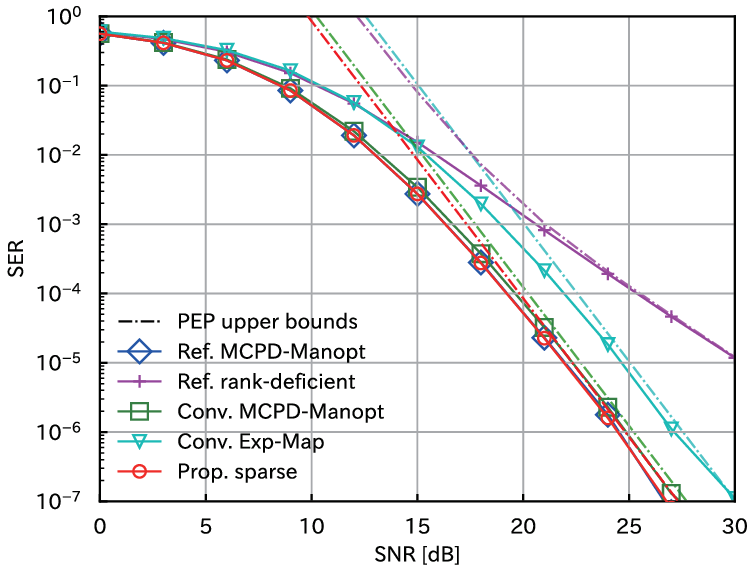}
    \caption{SER comparison with $(T,M)=(4,2)$ and $|\mathcal{X}|=4$.}
    \label{fig:SER_PEP_T=4_M=2_B=2}
\end{figure}

Fig.~\ref{fig:SER_PEP_T=4_M=2_B=2} shows SER comparisons and PEP evaluations for $(T,M)=(4,2)$, with the cardinality $|\mathcal{X}|=4$. Despite the constraint of sparse configuration, the proposed sparse constellation outperformed conventional methods such as Exp-Map and MCD-Manopt, and achieved performance that asymptotically approached the theoretically optimal constellation MCPD-Manopt. In the rank-deficient case, that is, when the rank of $\I_M-\X_i^\e\X_j\X_j^\e\X_i$ was deficient in the constellation as described in Section~\ref{sec:analysis}-A, the diversity gain decreased, and the slope of the SER curve was significantly smaller than other full-rank constellation. Furthermore, the proposed PEP accurately provided upper bounds for all constellations, regardless of rank deficiency. 

\begin{figure}[t]
    \centering
    \includegraphics[scale=0.7]{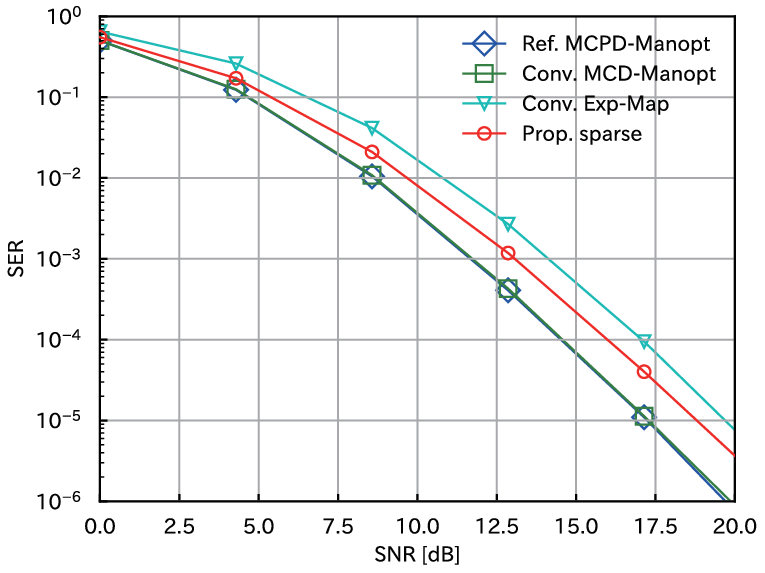}
    \caption{SER comparison with $(T,M)=(6,2)$ and $|\mathcal{X}|=64$.}
    \label{fig:SER_T=6_M=2_B=6}
\end{figure}

Fig.~\ref{fig:SER_T=6_M=2_B=6} compares SER performance for $(T,M)=(6,2)$ and $|\mathcal{X}|=64$. As shown in Fig.~\ref{fig:SER_T=6_M=2_B=6}, increasing the transmission rate, i.e., increasing the cardinality $|\mathcal{X}|$ prevented the proposed sparse constellation from asymptotically achieving the optimal performance. Although there existed an SNR loss compared with MCD-Manopt and MCPD-Manopt, the proposed constellation still outperformed Exp-Map and achieved suboptimal performance.

\begin{figure}[t]
    \centering
    \includegraphics[scale=0.7]{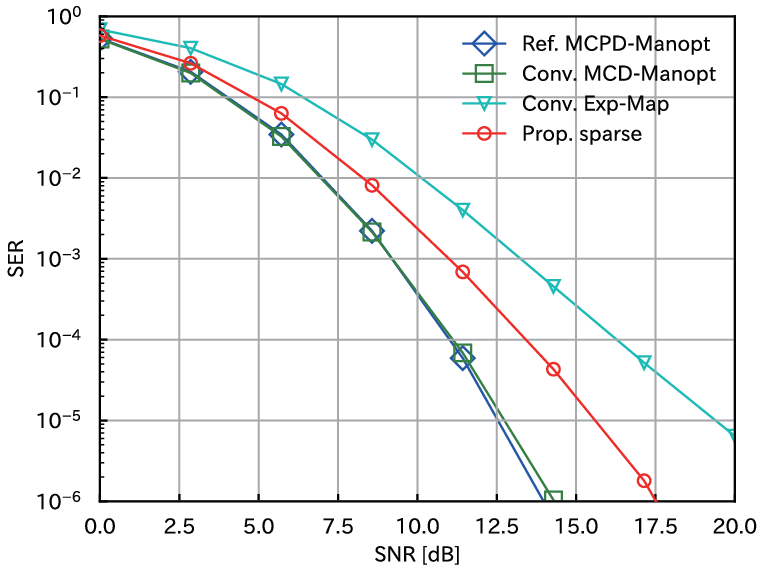}
    \caption{SER comparison with $(T,M)=(6,3)$ and $|\mathcal{X}|=64$.}
    \label{fig:SER_T=6_M=3_B=6}
\end{figure}

Fig.~\ref{fig:SER_T=6_M=3_B=6} presents the result for $(T,M)=(6,3)$ and $|\mathcal{X}|=64$.
Similar to Fig.~\ref{fig:SER_T=6_M=2_B=6}, although the performance loss became more significant as the cardinality increases, Fig.~\ref{fig:SER_T=6_M=3_B=6} demonstrates that the proposed sparse constellation still achieved superior performance compared with Exp-Map, even when the system parameters were varied.

\subsection{AMI Comparison}
Next, we compare AMI performance based on the calculation defined in~\eqref{eq:AMI}. The simulation settings were completely identical to those used in the SER comparison in Section~\ref{sec:res}-A.

\begin{figure}[tb]
    \centering
    \includegraphics[scale=0.7]{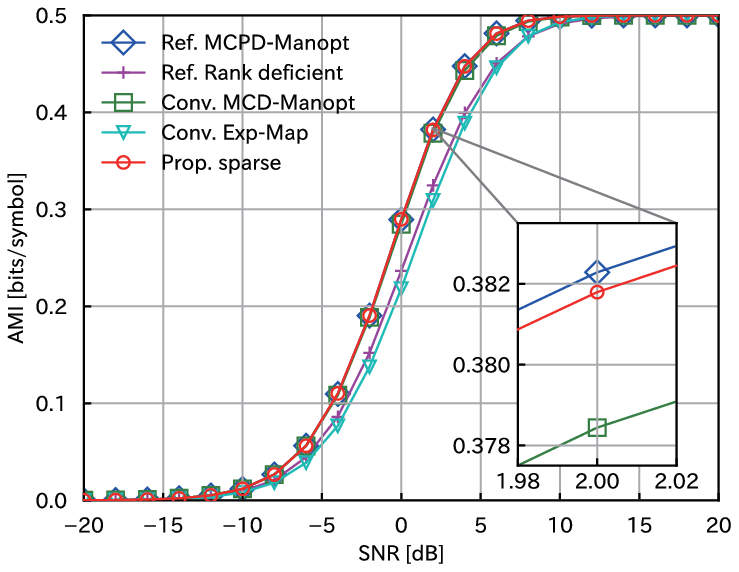}
    \caption{AMI comparison with $(T,M)=(4,2)$ and $|\mathcal{X}|=4$.}
    \label{fig:AMI_T=4_M=2_B=2}
\end{figure}

Fig.~\ref{fig:AMI_T=4_M=2_B=2} compares the AMI performance for $(T,M)=(4,2)$. Similar to the SER comparison in Fig.~\ref{fig:SER_PEP_T=4_M=2_B=2}, the proposed constellation outperformed the conventional constellations over the entire SNR range and exhibited performance asymptotically approaching that of MCPD-Manopt.
Furthermore, it is noteworthy that, as shown in Fig.~\ref{fig:AMI_T=4_M=2_B=2}, although the SER became worse than Exp-Map in the rank-deficient case, the relative performance relationship was reversed from the AMI perspective. This observation confirms that the maximization in~\eqref{eq:d_joint} effectively suppressed AMI degradation even under rank deficiency.


\begin{figure}[tb]
    \centering
    \includegraphics[scale=0.7]{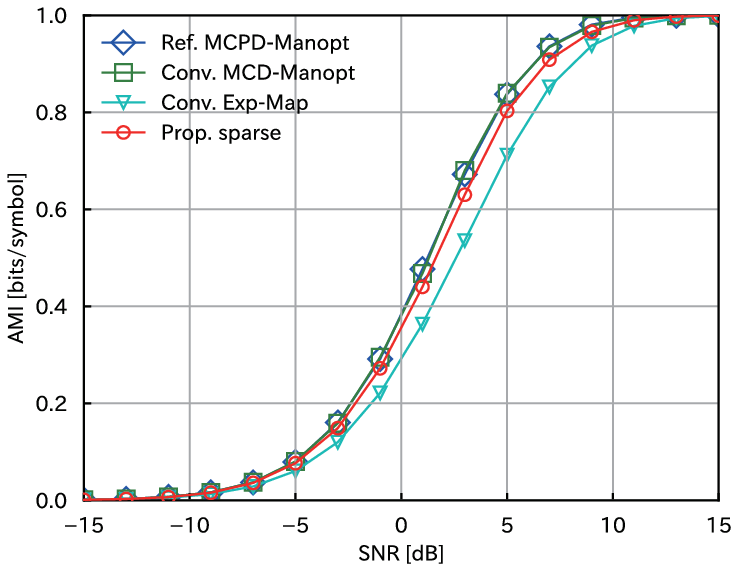}
    \caption{AMI comparison with $(T,M)=(6,2)$ and $|\mathcal{X}|=64$.}
    \label{fig:AMI_T=6_M=2_B=6}
\end{figure}

\begin{figure}[tb]
    \centering
    \includegraphics[scale=0.7]{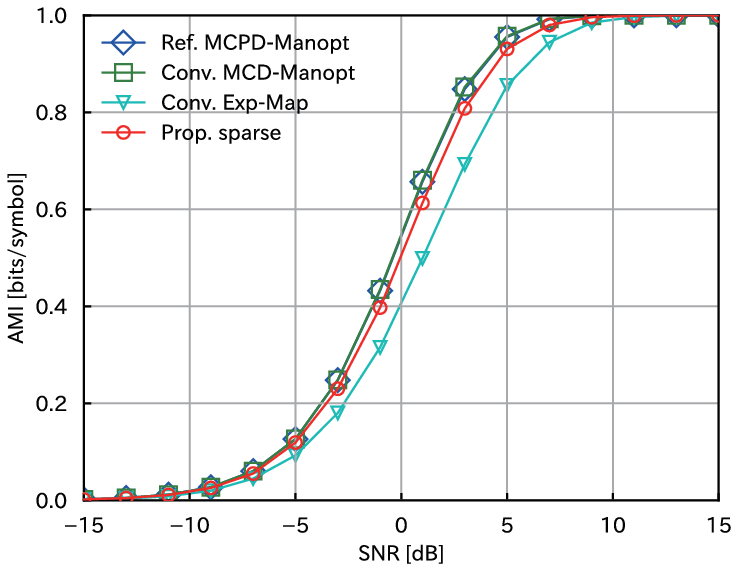}
    \caption{AMI comparison with $(T,M)=(6,3)$ and $|\mathcal{X}|=64$.}
    \label{fig:AMI_T=6_M=3_B=6}
\end{figure}

Figs.~\ref{fig:AMI_T=6_M=2_B=6} and~\ref{fig:AMI_T=6_M=3_B=6} compare the AMI performance for $(T,M)=(6,2)$ and $(6,3)$, respectively. In both cases, the proposed constellation outperformed Exp-Map and exhibited suboptimal performance compared with MCD-Manopt and MCPD-Manopt, although the advantage was less pronounced than in Fig.~\ref{fig:AMI_T=4_M=2_B=2}.

\subsection{Scalability}
\begin{figure*}[tb]
  \centering
  \subfloat[$(T,M)=(4,2)$]{%
    \includegraphics[width=0.33\linewidth]{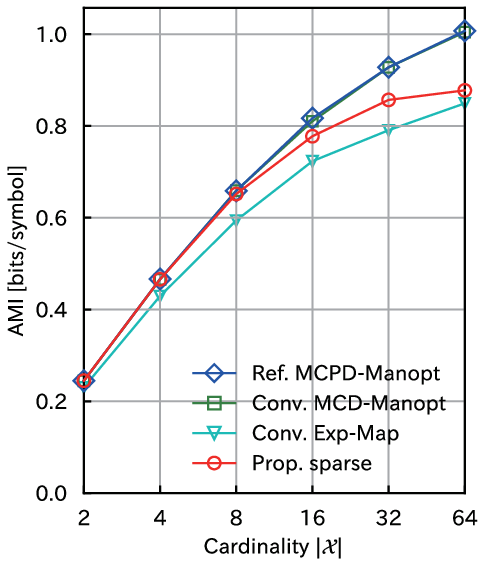}%
  }  
  \subfloat[$(T,M)=(6,2)$]{%
    \includegraphics[width=0.33\linewidth]{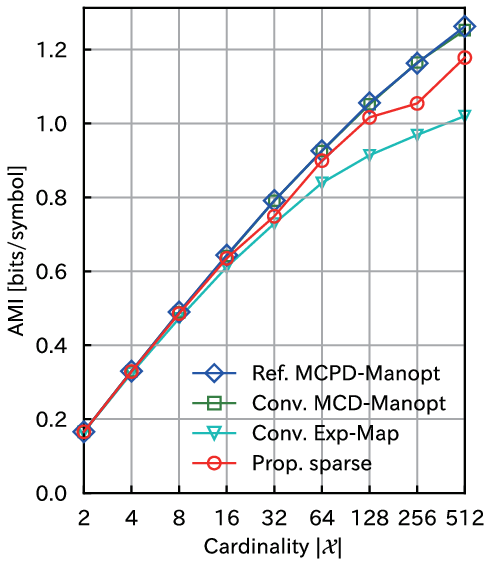}%
  }  
  \subfloat[$(T,M)=(6,3)$]{%
    \includegraphics[width=0.33\linewidth]{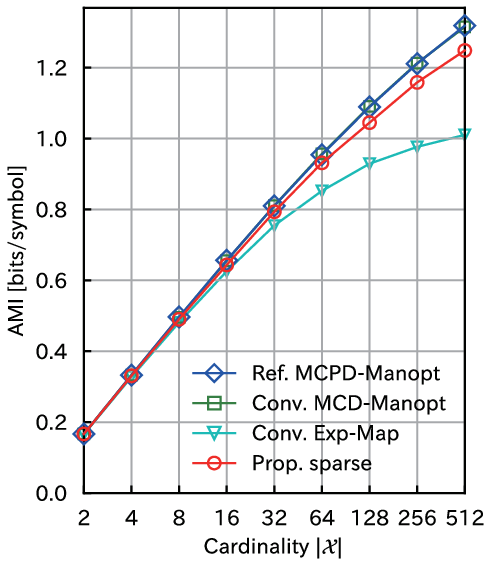}%
  }
  \caption{Comparison of AMI at different cardinality for $(T,M)=(4,2)$, $(6,2)$, and $(6,3)$.}
  \label{fig:scalability}
\end{figure*}

Fig.~\ref{fig:scalability} shows the AMI performance for $(T,M)=(4,2)$, $(6,2)$, and $(6,3)$ at different cardinalities $|\mathcal{X}|$. For all simulations, the SNR was fixed at 5 dB. As shown in Fig.~\ref{fig:scalability}, the proposed constellation outperformed Exp-Map for all $(T,M)$ settings. At low cardinalities, the proposed constellation exhibited performance that asymptotically approached that of MCD-Manopt and MCPD-Manopt. However, as clearly observed in Fig.~\ref{fig:scalability}(a), the performance degradation became more significant as the cardinality increases. This behavior was attributed to the reduction in the degrees of freedom and expressive subspaces caused by the sparsity constraint. This observation implies that the proposed sparse constellation offers a clear advantage in scenarios with relatively low transmission rates.

\subsection{Complexity Analysis}
When performing the GLRT detection defined in~\eqref{eq:GLRT}, the time complexity of exhaustive search for a general dense matrix is $\O(|\mathcal{X}|TMN)$, where $\X\in\C^{T\times M}$ and $\Y\in\C^{T\times N}$ denote the transmit and receive signal matrices, respectively, and $|\mathcal{X}|$ is the cardinality. In contrast, since the proposed sparse constellation contains at most one nonzero element in each row, the computational time complexity is linearly reduced to $\O(|\mathcal{X}|TN)$ under the same detection scheme. This independence from the number of transmit antennas $M$ makes the proposed method particularly effective for large-antenna configurations.
These complexity results are summarized in Table~\ref{tab:complexity}.

Furthermore, the proposed sparse constellation also reduces the space complexity. Sparse matrix storage schemes are summarized in~\cite{bell2008efficient}. Owing to the constraints defined in Section~\ref{sec:prop}, each row of the proposed constellation matrix stores only one nonzero element, and thus the ELLPACK format~\cite{kincaid1989ITPACKV} is most suitable for storage. For a general dense matrix with the same matrix dimensions and cardinality, the space complexity is $\O(|\mathcal{X}|TM)$. In contrast, for the proposed sparse constellation, employing the ELLPACK storage scheme reduces the space complexity to $\O(|\mathcal{X}|T)$, because only the positions of the nonzero elements and their complex values need to be stored. Hence, similar to the time complexity, the space complexity is also reduced to linear order with respect to $M$, as given in Table~\ref{tab:complexity}.

\begin{table}[tb]
    \centering
    \caption{Computational complexity comparison in GLRT detection.}
    \begin{tabular}{|c|c|c|} \hline
     & Conventional dense & Proposed sparse\\
     & constellation  & constellation\\\hline
      Time complexity   & $\mathcal{O}(|\mathcal{X}|TMN)$ & $\mathcal{O}(|\mathcal{X}|TN)$  \\ \hline
      Space complexity & $\mathcal{O}(|\mathcal{X}|TM)$ & $\mathcal{O}(|\mathcal{X}|T)$ \\ \hline
    \end{tabular}
    \label{tab:complexity}
\end{table}

\section{Conclusion}
\label{sec:conc}
In this paper, we proposed a sparse Grassmannian code design for noncoherent MIMO systems by exploiting the fact that the Schubert cell decomposition of the Grassmann manifold naturally induces sparsity. We modified the PEP formulation for noncoherent GLRT detection under uncorrelated Rayleigh fading channels, addressing the critical limitation that the conventional PEP does not provide an appropriate upper bound in rank-deficient cases of $\I_M-\X_i^\e\X_j\X_j^\e\X_i$. In addition, under the same detection scheme, we derived a closed-form metric that effectively maximizes the AMI for a given SNR and system configuration. Numerical results demonstrated that the proposed sparse constellation outperformed conventional constellations in both SER and AMI, and asymptotically approached the theoretically optimal performance in certain scenarios. Furthermore, the proposed constellation was shown to maintain effective performance over various parameter settings, while enabling a linear reduction in computational complexity for GLRT detection in terms of both time and space requirements. These results indicate that the proposed sparse Grassmannian constellation provides a practical and efficient solution for noncoherent MIMO communication.

\appendices
\section{Proof of Theorem~\ref{thm:conc}}
\label{app:conc}
In the inequality \eqref{ineq:AMI}, the dominant contribution is the expectation 
\begin{align}
 \E_{\H,\V}\left[\exp(-\alpha \operatorname{tr}(\Y_i\Y^\e_i\Deltab)) \right]:=\mathcal{E}_{i,j}
 \label{eq:eps},
\end{align}
with $\alpha$ defined in~\eqref{eq:alpha}.
To evaluate this expectation $\mathcal{E}_{i,j}$ in closed-form, we vectorize $\H$ and $\V$, and obtain $\mathbf{h}$ and $\mathbf{v}$ as $\mathbf{h} = \vec(\H) \in \C^{MN\times1}$ and $\mathbf{v} = \vec(\V) \in \C^{TN\times1}$
For $\mathbf{h}$ and $\mathbf{v}$, we define $\mathbf{z}$ as
\begin{align}
    \mathbf{z} = \begin{bmatrix}
        \mathbf{h} \\
        \mathbf{v}
    \end{bmatrix}
    \in\C^{(MN+TN)\times1}
    \sim\CN(\mathbf{0}, \mathbf{\Sigma})
    \label{eq:z},
\end{align}
where $\mathbf{\Sigma}=\operatorname{diag}(\I_{MN}, \sigma_v^2\I_{TN})\in\C^{(MN+TN)\times (MN+TN)}$. For $\Y_i=\X_i\H+\V$, applying the vector formula $\vec(\mathbf{ABC}) = (\mathbf{C}^\mathrm{T}\kr \mathbf{A})\vec(\mathbf{B})$, we obtain
\begin{align}
    \vec(\Y_i) &=(\I_{N}\kr\X_i)\mathbf{h}+\mathbf{v} = \mathbf{B}\mathbf{z} 
    \label{eq:vecY},
\end{align}
where $\mathbf{B}=[\I_N\kr\X_i,~\I_{TN}]\in\C^{TN\times(MN+TN)}$. Then, $\operatorname{tr}(\Y_i\Y^\e_i\Deltab)$ can be written in vector form as
\begin{align}
    \operatorname{tr}(\Y_i\Y^\e_i\Deltab) 
    &= \vec(\Y_i)^\e(\I_N\kr\Deltab)\vec(\Y_i) \notag \\ 
    & = \mathbf{z}^\e\mathbf{B}^\e(\I_N\kr\Deltab)\mathbf{B}\mathbf{z} \notag \\
    & = \mathbf{z}^\e\mathbf{A}\mathbf{z} 
    \label{eq:trY_i},
\end{align}
where $\mathbf{A}=\mathbf{B}^\e(\I_N\kr\Deltab)\mathbf{B}\in\C^{(MN+TN)\times(MN+TN)}$. The expectation $\mathcal{E}_{i,j}$ can be written using \eqref{eq:trY_i}, as
\begin{align}
     \mathcal{E}_{i,j}  
     & = \E_{\mathbf{z}}[\exp(-\alpha\cdot\mathbf{z}^\e\mathbf{A}\mathbf{z})].
     \label{eq:E}
\end{align}
Since $\mathbf{z}$ follows the complex Gaussian distribution with the covariance matrix $\mathbf{\Sigma}$, we have~\cite{goodman1963Statistical}
\begin{align}
    \E_{\mathbf{z}}[\exp(-\alpha\cdot\mathbf{z}^\e\mathbf{A}\mathbf{z})]=\det(\I_{MN+TN}+\alpha\mathbf{\Sigma}\A)^{-1}
    \label{eq:Ez},
\end{align}
while supporting any value of $T,M,$ and $\sigma_v^2>0$.
The matrix $\I_{MN+TN}+\alpha\mathbf{\Sigma}\A$ can be partitioned into four blocks as
\begin{align}
    \I_{MN+TN}+\alpha\mathbf{\Sigma}\A = \begin{bmatrix}
        \I_{MN}+\alpha\A_{11} & \alpha\A_{12} \\
        \alpha\sigma_v^2\A_{21} & \I_{TN}+\alpha\sigma_v^2\A_{22}
    \end{bmatrix}
    \label{eq:blcokdA},
\end{align}
where the sub-matrices $\A_{11}, \A_{12}, \A_{21}, \A_{22}$ are defined as
\begin{align}
    \A_{11} 
    &= \I_N\kr(\X_i^\e\Deltab\X_i)\in\C^{MN\times MN}
    \label{eq:A11},\\
    \A_{12} 
    &= \I_N\kr(\X_i^\e\Deltab)\in\C^{MN\times TN}
    \label{eq:A12},\\
    \A_{21} 
    &= \I_N\kr(\Deltab\X_i) \in\C^{TN\times MN}
    \label{eq:A21},\\
    \A_{22} 
    &= \I_N\kr \Deltab\in\C^{TN\times TN}
    \label{eq:A22}.
\end{align}
Applying the formula for the determinant of a block matrix, the determinant $\det(\I_{MN+TN}+\alpha\mathbf{\Sigma}\A)$ can be rewritten as
\begin{align}
    &\det(\I_{MN+TN}+\alpha\mathbf{\Sigma}\A) 
    := D_1\cdot D_2
    \label{eq:blockdet},
\end{align}
where $D_1=\det(\I_{TN}+\alpha\sigma_v^2\A_{22})$ and $D_2=\det(\I_{MN}+\alpha\A_{11}-\alpha\A_{12}(\I_{TN}+\alpha\sigma_v^2\A_{22})^{-1}\alpha\sigma_v^2\A_{21})$. Here, these determinants can be expressed in terms of the principal angles between $\X_i$ and $\X_j$ defined in~\eqref{eq:p_angle}, as stated below.
\begin{lem}
\label{lem:D1}
The determinant $D_1$ can be expressed in terms of the principal angles as
\begin{align}
    D_1
    & = \left[\prod_{m=1}^M \left(1-\alpha^2\sigma_v^4\sin^{2}\theta_m^{(i,j)}\right) \right]^N
    \label{eq:D_1_p}.
\end{align}
\end{lem}
The proof is given in Appendix~\ref{app:D1}. 
\begin{lem}
\label{lem:D2}
Similarly, the determinant $D_2$ can be expressed in terms of the principal angles as
\begin{align}
    D_2 
    &= \left[\prod_{m=1}^{M}\left(\frac{1+(\alpha-\alpha^2\sigma_v^2-\alpha^2\sigma_v^4)\sin^2\theta_m^{(i,j)}}{1-\alpha^2\sigma_v^4\sin^2\theta_m^{(i,j)}}\right)\right]^N
    \label{eq:D2_p}.
\end{align}
\end{lem}
The proof is given in Appendix~\ref{app:D2}. 
With these results, we now derive a closed-form expression for $\mathcal{E}_{i,j}$ as
\begin{align}
    \mathcal{E}_{i,j} &= \det(\I_{MN+TN}+\alpha\mathbf{\Sigma}\A)^{-1} \notag \\
    &= (D_1D_2)^{-1} \notag \\
    &= \left[\prod_{m=1}^{M} {\left(1+(\alpha-\alpha^2\sigma_v^2-\alpha^2\sigma_v^4)\sin^2\theta_m^{(i,j)}\right)} \right]^{-N} \notag \\ 
    & = \left[\prod_{m=1}^{M} {\left(1+\kappa(\lambda)\sin^2\theta_m^{(i,j)}\right)} \right]^{-N}.
    \label{eq:resE}
\end{align}

\section{Proof of Lemma~\ref{lem:D1}}
\label{app:D1}
First, we use the fact that the eigenvalues of the matrix $\Deltab$ are denoted as $\pm \sin\theta_1^{(i,j)},\cdots, \pm\sin\theta_M^{(i,j)}$.
Let $\u_m \in \mathrm{col}(\X_i)$, where $\mathrm{col}(\X_i)$ denotes a linear subspace spanned by all column vectors of $\X_i$, i.e., $\mathrm{col}(\X_i)=\mathrm{span}\{\x_1,\cdots, \x_M \}$ where $\x_j$ is a column vector of $j$-th column of $\X_i$. In this setting, $\u_m$ can be regarded as a representative vector on the $\X_i$ of the subspace that defines the $m$-th principal angle $\theta^{(i,j)}_m$. Now, let $\mathbf{v}_m$ be a unit-norm column vector that is orthogonal to $\u_m$, that is, $\u_m\perp\mathbf{v}_m$ and $\|\mathbf{v}_m\|=1$. Since $\u_m$ and $\mathbf{v}_m$ are mutually orthogonal, they span a two-dimensional space. Denoting this subspace $S_m$, we define
\begin{align}
    S_m=\text{span}\{\mathbf{u}_m,\mathbf{v}_m\}
    \label{eq:Sm}.
\end{align}
Within this two-dimensional space, the vector $\w_m$ defined as
\begin{align}
    \mathbf{w}_m=\cos\theta_m^{(i,j)}\mathbf{u}_m+\sin\theta_m^{(i,j)}\mathbf{v}_m
    \label{eq:wm}
\end{align}
is obtained by rotating $\u_m$ by angle $\theta_m^{(i,j)}$ within the two-dimensional space $S_m$, and satisfies $\w_m\in\text{col}(\X_j)$. Let $\Deltab_m$ denote the matrix obtained by projecting the matrix $\Deltab$ onto the previously defined two-dimensional space $S_m$. Since the projection matrices onto $\text{col}(\X_i)$ and $\text{col}(\X_j)$ are given by the projection matrices $\P_i$ and $\P_j$ respectively, we have $\Deltab=\P_i-\P_j$. Thus, it is sufficient to consider the projections of $\P_i$ and $\P_j$ onto $S_m$. The projection matrix $\P_i$ orthogonally projects an arbitrary column vector onto $\text{col}(\X_i)$. From the previous definitions, using the facts that $\u_m\in\col(\X_i)$ and $\mathbf{v}_m\perp\col(\X_i)$, we obtain $\P_i\u_m=\u_m$ and $\P_i\mathbf{v}_m=\mathbf{0}$.
Let $\P_{i,m}$ denote the matrix representation of $\P_i$ after projection onto $S_m$, which is written as
\begin{align}
    \P_{i,m}=\begin{bmatrix}
        \mathbf{u}_m^\e\P_i\mathbf{u}_m & \mathbf{u}_m^\e\P_i\mathbf{v}_m \\
        \mathbf{v}_m^\e\P_i\mathbf{u}_m & \mathbf{v}_m^\e\P_i\mathbf{v}_m 
    \end{bmatrix}
    =\begin{bmatrix}
        1 & 0 \\
        0 & 0
    \end{bmatrix},
    \label{eq:P_im}
\end{align}
where we use the fact that $\u_m^\e\u_m=1$. Similarly, since $\w_m\in\col(\X_j)$, we have $\P_j\u_m=\cos\theta_m^{(i,j)}\w_m$ and $\P_j\mathbf{v}_m=\sin\theta_m^{(i,j)}\w_m$. Hence, the matrix representation of $\P_j$ projected onto $S_m$ denoted as $\P_{j,m}$ is given by 
\begin{align}
    \P_{j,m}&=\begin{bmatrix}
        \mathbf{u}_m^\e\P_j\mathbf{u}_m & \mathbf{u}_m^\e\P_j\mathbf{v}_m \\
        \mathbf{v}_m^\e\P_j\mathbf{u}_m & \mathbf{v}_m^\e\P_j\mathbf{v}_m 
    \end{bmatrix} \notag \\
    &=\begin{bmatrix}
        \mathbf{u}_m^\e\cos\theta_m^{(i,j)}\mathbf{w}_m & \mathbf{u}_m^\e\sin\theta_m^{(i,j)}\mathbf{w}_m \\
        \mathbf{v}_m^\e\cos\theta_m^{(i,j)}\mathbf{w}_m & \mathbf{v}_m^\e\sin\theta_m^{(i,j)}\mathbf{w}_m
    \end{bmatrix}
\notag \\
    & = \begin{bmatrix}
        \cos^2\theta_m^{(i,j)} & \sin\theta_m^{(i,j)}\cos\theta_m^{(i,j)} \\
        \sin\theta_m^{(i,j)}\cos\theta_m^{(i,j)} & \sin^2\theta_m^{(i,j)}
    \end{bmatrix}    
\label{eq:P_jm},
\end{align}
where we use the fact that $\mathbf{v}^\e\mathbf{v}=1$. From \eqref{eq:P_im} and \eqref{eq:P_jm}, $\Deltab_m$ can be expressed solely in terms of the principal angles as
\begin{align}
    \Deltab_m&=\P_{i,m}-\P_{j,m} \notag \\
    &=\begin{bmatrix}
        1-\cos^2\theta_m^{(i,j)} & -\sin\theta_m^{(i,j)}\cos\theta_m^{(i,j)} \\
        -\sin\theta_m^{(i,j)}\cos\theta_m^{(i,j)} & -\sin^2\theta_m^{(i,j)}
    \end{bmatrix}
    \notag \\
    & = \begin{bmatrix}
        \sin^2\theta_m^{(i,j)} & -\sin\theta_m^{(i,j)}\cos\theta_m^{(i,j)} \\
        -\sin\theta_m^{(i,j)}\cos\theta_m^{(i,j)} & -\sin^2\theta_m^{(i,j)}
    \end{bmatrix}.
    \label{eq:Deltab_m}
\end{align}
The  eigenvalues of $\Deltab_m$ can be denoted from the result of~\eqref{eq:Deltab_m} as $\pm\sin\theta_m^{(i,j)}$. The  eigenvalues of $\Deltab$ can be calculated by the result of $\Deltab_m$ for $M$ principal angles, which is denoted as $\pm \sin\theta_1^{(i,j)},\cdots, \pm\sin\theta_M^{(i,j)}$ 

From \eqref{eq:A22}, $D_1$ is expressed by principal angles as
\begin{align}
    D_1
    &=\det(\I_{TN}+\alpha\sigma_v^2\A_{22}) \notag \\
    &= \det(\I_{TN}+\alpha\sigma_v^2(\I_N\kr \Deltab)) \notag \\
    & = \det(\I_T+\alpha\sigma_v^2\Deltab)^N \notag \\    
    & = \left[\prod_{m=1}^M \left(1-\alpha^2\sigma_v^4\sin^{2}\theta_m^{(i,j)}\right) \right]^N
    \label{eq:D_1_p_2}.
\end{align}

\section{Proof of Lemma~\ref{lem:D2}}
\label{app:D2}
Compared to $D_1$, $D_2$ is difficult to express simply because it includes the inverse matrix term. Hence, we simplify the evaluation by decomposing the problem into two-dimensional subspaces associated with each principal angle. First, we consider expressing it as a matrix product without using the Kronecker product.
From \eqref{eq:A11} to \eqref{eq:A22}, we have
\begin{align}
    &\alpha\A_{12}(\I_{TN}+\alpha\sigma_v^2\A_{22})^{-1}\alpha\sigma_v^2\A_{21} \notag \\
    =& \I_N\kr(\alpha^2\X_i^\e\Deltab(\I_T+\alpha\sigma_v^2\Deltab)^{-1}\sigma_v^2\Deltab\X_i).
\end{align}
Thus, we obtain
\begin{align}
    &\I_{MN}+\alpha\A_{11}-\alpha\A_{12}(\I_{TN}+\alpha\sigma_v^2\A_{22})^{-1}\alpha\sigma_v^2\A_{21} \notag \\ 
    =& \I_{MN}+\I_N\kr(\alpha\X_i^\e\Deltab\X_i) \notag \\
    &-\I_N\kr\left(\alpha^2\X_i^\e\Deltab(\I_T+\alpha\sigma_v^2\Deltab)^{-1}\sigma_v^2\Deltab\X_i\right) \notag \\
    =&\I_N\kr \left(\I_M+\alpha\X_i^\e\Deltab(\I_T+\alpha\sigma_v^2\Deltab)^{-1}\X_i\right)
    \label{eq:d}.
\end{align}
Then $D_2$ can be denoted without using the Kronecker product as
\begin{align}
    D_2 
    & = \det \left( \I_N\kr(\I_M+\alpha\X_i^\e\Deltab(\I_T+\alpha\sigma_v^2\Deltab)^{-1}\X_i)\right) \notag \\
    &= \det \left(\I_M+\alpha\X_i^\e\Deltab(\I_T+\alpha\sigma_v^2\Deltab)^{-1}\X_i \right)^{N}.
    \label{eq:D2}
\end{align}
Letting $\K =\alpha \X_i^\e\Deltab(\I_T+\alpha\sigma_v^2\Deltab)^{-1}\X_i$, we now consider expressing $D_2 = \det(\I_M+\K)^N$ in terms of the principal angles. 
By using $\Deltab_m$ denoted in~\eqref{eq:Deltab_m} of Lemma~\ref{lem:D1}, the matrix obtained by projecting the inverse term $(\I_T+\alpha\sigma_v^2\Deltab)^{-1}$ in $\K$ onto $S_m$, namely $(\I_2 + \alpha\sigma_v^2\Deltab_m)^{-1}$ can be expressed as
\begin{align}
    &(\I_2 + \alpha\sigma_v^2\Deltab_m)^{-1} \notag \\
    =& \begin{bmatrix}
        1 + \alpha\sigma_v^2\sin^2\theta_m^{(i,j)} & -\alpha\sigma_v^2\cos\theta_m^{(i,j)}\sin\theta_m^{(i,j)} \\
        -\alpha\sigma_v^2\cos\theta_m^{(i,j)}\sin\theta_m^{(i,j)} & 1 -\alpha\sigma_v^2\sin^2\theta_m^{(i,j)}
    \end{bmatrix}^{-1} \notag \\
    =&\frac{1}{\gamma}
    \begin{bmatrix}
        1 - \alpha\sigma_v^2\sin^2\theta_m^{(i,j)} & \alpha\sigma_v^2\cos\theta_m^{(i,j)}\sin\theta_m^{(i,j)} \\
        \alpha\sigma_v^2\cos\theta_m^{(i,j)}\sin\theta_m^{(i,j)} & 1 +\alpha\sigma_v^2\sin^2\theta_m^{(i,j)}
    \end{bmatrix},
    \label{eq:detDm}
\end{align}
where $\gamma=1-\alpha^2\sigma_v^4\sin^2\theta_m^{(i,j)}$.

Moreover, the $(m,l)$-th element of $\K$, denoted by $\K_{(m,\ell)}$ can be written using $\u_m,\u_\ell\in\col(\X_i)$ as
\begin{align}
    \K_{(m,\ell)} = \alpha\u_m^{\e}\Deltab(\I_T+\alpha\sigma_v^2\Deltab)^{-1}\u_\ell
    \label{eq:K_ml}
\end{align}
Since $\u_m,\u_\ell\in\col(\X_i)$ and $\X_i$ represents a point on the Grassmann manifold $\G(T,M)$, its column vectors form an orthonormal set, thus
\begin{align}
    \mathbf{u}_m^\e\mathbf{u}_\ell=\left\{\begin{array}{cc}
        1 & (m=\ell) \\
        0 & (m\neq\ell)
    \end{array} \right.
    \label{eq:uml}
\end{align}
holds. Consequently, when $m\neq\ell$, we have $\K_{(m,\ell)}=0$, which implies that $\K$ is a diagonal matrix. For the diagonal element $\K_{(m,m)}$, the corresponding computation is equivalent to the expression in the space obtained by projecting onto the previously defined two-dimensional space $S_m$. Hence, by using the result in~\eqref{eq:detDm}, we obtain
\begin{align}
    \K_{(m,m)}&=\alpha\mathbf{u}_m^\e\Deltab(\I_T+\alpha\sigma_v^2\Deltab)^{-1}\mathbf{u}_m \notag \\
    &=\alpha(\mathbf{E}_m^\e\mathbf{u}_m)^\e\Deltab_m(\I_2+\alpha\sigma_v^2\Deltab_m)^{-1}(\mathbf{E}_m^\e\mathbf{u}_m),
\end{align}
where $\mathbf{E}_m \in\C^{T\times2}$ is the basis transformation matrix for the subspace $S_m$, which is defined as $\mathbf{E}_m=[\u_m~\mathbf{v}_m]$. Thus, since $\mathbf{E}_m^\e\u_m=[1,0]^\mathrm{T}$, the diagonal element $\K_{(m,m)}$ can be rewritten as
\begin{align}
    &\K_{(m,m)} \notag \\ =&\alpha(\mathbf{E}_m^\e\mathbf{u}_m)^\e\Deltab_m(\I_2+\alpha\sigma_v^2\Deltab_m)^{-1}(\mathbf{E}_m^\e\mathbf{u}_m) \notag \\
    =& \alpha[1,0]\Deltab_m(\I_2+\alpha\sigma_v^2\Deltab_m)^{-1}\begin{bmatrix}
        1 \\ 0
    \end{bmatrix} \notag \\
    =&\begin{bmatrix}
        \alpha \sin^2\theta_m^{(i,j)} & -\alpha \cos\theta_m^{(i,j)}\sin\theta_m^{(i,j)} \\
        -\alpha \cos\theta_m^{(i,j)}\sin\theta_m^{(i,j)} & -\alpha \sin^2\theta_m^{(i,j)}
    \end{bmatrix} 
    \notag \\
    \cdot& \frac{1}{\gamma} 
    \begin{bmatrix}
        1 - \alpha \sigma_v^2\sin^2\theta_m^{(i,j)} & \alpha \sigma_v^2\cos\theta_m^{(i,j)}\sin\theta_m^{(i,j)} \\
        \alpha \sigma_v^2\cos\theta_m^{(i,j)}\sin\theta_m^{(i,j)} & 1 +\alpha \sigma_v^2\sin^2\theta_m^{(i,j)}
    \end{bmatrix}_{(1,1)} \notag \\
    =& \frac{\alpha(1-\alpha\sigma_v^2)\sin^2\theta_m^{(i,j)}}{\gamma} 
    \label{eq:Kmm}.
\end{align}
Since $\K = \mathrm{diag}(\K_{(1,1)}, \cdots, \K_{(M,M)})$, $D_2$ finally can be expressed in terms of the principal angles as
\begin{align}
    D_2 &=\det(\I_M+\K)^N \notag \\
    &= \left[\prod_{m=1}^{M}(1+\K_{(m,m)})\right]^N \notag \\
    &= \left[\prod_{m=1}^{M}\left(\frac{1+(\alpha-\alpha^2\sigma_v^2-\alpha^2\sigma_v^4)\sin^2\theta_m^{(i,j)}}{1-\alpha^2\sigma_v^4\sin^2\theta_m^{(i,j)}}\right)\right]^N
    \label{eq:D2_p_2}.
\end{align}

\footnotesize{
	\bibliographystyle{IEEEtranURLandMonthDiactivated}
	\bibliography{main}
}

\end{document}